\documentstyle[epsfig,12pt,preprint,tighten,aps]{revtex}
\begin{document}

\draft

\title{\rightline{{\tt (April 1998)}}
\rightline{{\tt UM-P-98/17}}
\rightline{{\tt RCHEP-98/05}}
\ \\
Relic neutrino asymmetries and big bang nucleosynthesis in 
a four neutrino model}
\author{N. F. Bell, R. Foot and R. R. Volkas }
\address{School of Physics\\
Research Centre for High Energy Physics\\
The University of Melbourne\\
Parkville 3052 Australia\\
(n.bell@physics.unimelb.edu.au, foot@physics.unimelb.edu.au, 
r.volkas@physics.unimelb.edu.au)}
\maketitle

\begin{abstract}

Oscillations between ordinary and sterile neutrinos 
can generate large neutrino
asymmetries in the early universe.  These asymmetries can 
significantly affect big
bang nucleosynthesis (BBN) through modification of 
nuclear reaction rates.  We study
this phenomenon within a model consisting of 
the three ordinary neutrinos plus one
sterile neutrino that can be motivated by the 
neutrino anomalies and the dark matter
problem.  We calculate how the lepton asymmetries 
produced evolve at temperatures
where they impact on BBN.  The effect of 
the asymmetries on primordial helium
production is determined, leading to an effective 
number of neutrino flavours during
BBN of either about $2.7$ or $3.1$ depending 
on the sign of the lepton asymmetry.

\end{abstract}

\section{Introduction}

Models with sterile neutrinos are currently 
favoured to accommodate all the neutrino
oscillation measurements.  It is difficult to 
simultaneously account for the solar,
atmospheric and LSND anomalies with just three ordinary 
neutrinos and hence only two
mass-squared differences.  Therefore we require at least 
one additional neutrino if we wish to explain all
three of the observed neutrino anomalies.
Constraints from the Z-width suggest that any additional
neutrinos must be sterile.
Oscillations between ordinary and sterile
neutrinos in the early universe have been studied in
\cite{enqvist,early,kirilova,rr1,shi} and found to generate 
large neutrino asymmetries\cite{rr1,shi}, 
which can significantly affect big bang nucleosynthesis
(BBN) reaction rates, and can result 
in an effective neutrino number of less than
three during BBN \cite{rr}. In this paper, the 
techniques developed in Ref.\cite{rr}
will be applied to the four neutrino model 
of Ref.\cite{model} which has almost degenerate 
$\nu_\mu$ and $\nu_\tau$. This model is distinct
from the four neutrino
model of Ref.\cite{rr} where a mass hierarchy between $\nu_{\mu}$ and 
$\nu_{\tau}$ was assumed.

A motivation for studying how neutrino oscillations can 
affect $N_{\nu}^{eff}$, the
effective neutrino number during nucleosynthesis, 
arises from discrepancies between
the observed deuterium abundance and standard BBN.  
The primordial D/H ratio can be used
to give a sensitive determination of the baryon 
to photon ratio $\eta$ which, given
the inferred primordial $^4$He mass fraction, can be 
used to predict the number of light
neutrinos $N_{\nu}^{eff}$.

Although there are numerous measurements of the 
deuterium abundance, obtaining an
upper bound to the abundance is quite difficult, 
and there are conflicting
observations.  A high deuterium measurement of 
$D/H=(1.9\pm0.4)\times10^{-4}$ was
found in Ref.\cite{rugers}, leading 
to $\eta\sim2\times10^{-10}$, while in
Ref.\cite{tytler} a low deuterium result 
of $D/H=(2.3\pm0.3\pm0.3)\times10^{-5}$ was
obtained, suggesting that $\eta\sim7\times10^{-10}$.  From 
these values of $\eta$, and the helium mass fraction $Y_P$, 
the effective number of neutrinos can be calculated.  By way 
of example, the results
\begin{eqnarray}
N_{\nu}^{eff}=2.9\pm0.3  \text{   high D/H},  \nonumber \\
N_{\nu}^{eff}=1.9\pm0.3  \text{   low D/H},
\end{eqnarray}
were obtained in Ref.\cite{hata}.  (For other calculations 
of $N^{eff}_{\nu}$ see Ref.\cite{others}.)
Hence, if the low deuterium result is correct, standard BBN 
would require an effective
neutrino number of less than three (provided
that the value of $Y_P$
used in Ref.\cite{hata} is correct). 
Clearly $N_{\nu}^{eff} < 3$ would imply new physics.
Further observations are required to pin down the D/H value (and
also $Y_P$)
although it now seems as though the low D/H 
result is favoured \cite{tytler,turner}. 

One possibility for the new physics is to look at the
neutrino sector and study mechanisms
by which the effective number of neutrinos can be 
made less than three.  A possible
way of doing this is to create an electron-neutrino asymmetry, 
as this would directly
affect the reaction rates which determine the n/p 
ratio just before nucleosynthesis. 
The n/p ratio controls the helium mass fraction $Y_P$ via the equation
\begin{equation}
\frac{dY_P}{dt}=-\lambda(n\rightarrow p)Y_P+\lambda(p\rightarrow n)(2-Y_P),
\end{equation}
where the reaction rates
\begin{eqnarray}
\label{eq:rates}
\lambda(n\rightarrow p)\simeq \lambda(n + \nu_e \rightarrow p + e^-) 
+ \lambda(n + e^+ \rightarrow p + \bar{\nu}_e), \nonumber \\
\lambda(p\rightarrow n) \simeq \lambda(p + e^- \rightarrow n + \nu_e) 
+ \lambda(p + \bar{\nu}_e \rightarrow n + e^+),
\end{eqnarray}
depend on the momentum distributions of the species involved.  
The processes in Eq.(\ref{eq:rates}) for 
determining $n\leftrightarrow p$ are valid
for temperatures above about 0.4 MeV, below which the 
weak interaction rates become
frozen out and neutron decay becomes the dominant 
factor affecting the n/p ratio.  An
excess of $\nu_e$ over $\bar{\nu}_e$ would reduce the n/p 
ratio, due to the modification of the neutrino momentum 
distributions through non-zero chemical
potentials, thereby changing the rates for the processes in 
Eq.(\ref{eq:rates}). 
Neutron decay is not significantly altered by 
lepton asymmetries.  A change in the
helium mass fraction $Y_P$ can be related to a 
change in the effective number of
neutrinos by \cite{walker}
\begin{equation}
\delta Y_P \simeq 0.012\times \delta N_{\nu}^{eff}.
\end{equation}
A large neutrino asymmetry could simply 
be postulated, but it is more interesting to
look at physics of the neutrino sector which 
will allow a large neutrino asymmetry to
evolve as a dynamical variable. 

A four-neutrino model was studied in Ref.\cite{rr}, involving 
the three ordinary neutrinos and one sterile neutrino, 
where the mass hierarchy $m_{\nu_{\tau}}\gg
m_{\nu_{\mu}}, m_{\nu_e}, m_{\nu_s}$ was 
considered by way of example.  In
this case oscillations between
$\bar{\nu}_{\tau}$ and $\bar{\nu}_s$ resulted in 
an excess of $\nu_{\tau}$ over
$\bar{\nu}_{\tau}$ thereby generating a large tau-neutrino 
asymmetry and hence a
large tau-lepton number.  $\bar{\nu}_{\tau}-\bar{\nu}_e$ 
oscillations also occurred,
and it was shown that this allowed some of the tau-neutrino 
asymmetry to be transferred to an electron-neutrino asymmetry. 
The effective number of
neutrinos found in Ref.\cite{rr} was either 2.5 or 3.4.  
The two values arise because
there is an ambiguity involving the sign of 
the asymmetry and hence the prediction
for $N^{eff}_{\nu}$ (see Ref.\cite{rr1} for 
a discussion of this issue).  For a
positive asymmetry, $\delta N_{\nu}^{eff} \simeq -0.5$ 
was obtained over a range of
mass differences $\delta m_{\tau s}^2\sim 10-3000\text{eV}^2$, 
while for a negative
asymmetry the result was $\delta N_{\nu}^{eff} \simeq +0.4$.  
So we have the interesting situation that $N_{\nu}^{eff}<3$ 
can be achieved although the precise value of $N_{\nu}^{eff}$
is model dependent as this paper will illustrate.
If future measurements can pin down $N_{\nu}^{eff}$ precisely
enough, then this quantity can be used to help discriminate
various competing models explaining the observed neutrino
physics anomalies.

\section{Model of neutrino masses}

The model we will consider in this paper is again a system 
of four neutrinos:
$\nu_\tau$, $\nu_\mu$, $\nu_e$ and $\nu_s$. In this
model the $\nu_e$ and $\nu_s$ are assumed to 
be very light $(\ll1\text{eV})$, and
$\nu_\tau$ and $\nu_ \mu$ are taken to have 
nearly degenerate masses.  This 
model has been proposed in \cite{model} as a pattern
of masses that will allow the data from the solar, 
atmospheric, and LSND experiments
to be satisfied as well as being consistent with dark matter 
models.  

In this scenario, the solar neutrino problem is 
explained by the small angle MSW
solution with oscillations between $\nu_e$ and $\nu_s$ 
with $\delta m_{es}^2\sim 10^{-5}\text{eV}^2$ and 
$\sin^22\theta \sim 10^{-2}$\cite{b}, while the atmospheric neutrino
deficit is explained by $\nu_{\tau}-\nu_{\mu}$ oscillations\cite{b2} 
with $\delta m_{\tau\mu}^2\sim10^{-2}-10^{-3}\text{eV}^2$ 
and $\sin^22\theta \simeq 1$.  The LSND data
requires a $\nu_\mu-\nu_e$ solution with 
$\delta m_{\mu e}^2$ in the range $0.2 -
10\text{eV}^2$ and 
$\sin^22\theta \simeq 3\times 10^{-2}-10^{-3}$ \cite{lsnd}. 
It has been argued\cite{primack} from considerations
of structure formation in the early Universe that the optimal value
for the degenerate $\nu_\mu$ and $\nu_\tau$ masses is $2.4$eV.
Note, however, that the creation of a large $L_{\mu}$ and $L_{\tau}$
neutrino asymmetry should impact on the favoured 
neutrino mass for structure formation and dark matter.
For this reason it may well be necessary for
the optimal mass of approximately 2.4 eV to 
be recalculated, taking into account the modified neutrino 
number densities.

\section{Neutrino asymmetries and consequences for BBN}

For oscillations between $\nu_{\alpha}$ and $\nu_s$, 
the weak eigenstates are linear
combinations of the mass eigenstates 
$\nu_1$ and $\nu_2$,
\begin{eqnarray}
\nu_{\alpha}=\cos\theta_0\nu_1 + \sin\theta_0\nu_2, \nonumber \\
\nu_s=-\sin\theta_0\nu_1 + \cos\theta_0\nu_2,
\end{eqnarray}
where $\theta_0$ is the vacuum mixing angle (we assume that
$\theta_0$ is defined such that $\cos 2\theta_0 > 0$).
The mass-squared difference between the
two eigenstates is defined as 
$\delta m^2_{\alpha s}=m^2_2-m^2_1$.
The neutrinos
interact with matter, so a matter mixing angle 
$\theta_m$ is defined, which is
related to $\theta_0$ by \cite{kuo},
\begin{equation}
\sin^22\theta_m = \frac{\sin^22\theta_0}{\sin^22\theta_0+
(b \pm a -\cos2\theta_0)^2}.
\end{equation}
The term $(b \pm a)$ is related to the effective potential 
due to interactions of the
neutrinos with matter: 
\begin{equation}
V=\frac{\delta m^2_{\alpha s}}{2p}(b\pm a),
\end{equation}
where -/+ corresponds to $\nu/\bar{\nu}$ and $p$ 
is the neutrino momentum.
The functions $a$ and $b$ are given by
\begin{eqnarray}
\label{eq:a}
a=-\frac{4\sqrt{2}\zeta(3)G_FT^3L^{(\alpha)}p}
{\pi^2\delta m^2_{\alpha s}}, \nonumber \\
b=-\frac{4\sqrt{2}\zeta(3)G_FT^4A_{\alpha}p^2}
{\pi^2\delta m^2_{\alpha s}M_W^2},
\end{eqnarray}
where $A_e\simeq17$ and $A_{\tau,\mu}\simeq 4.9$\cite{rn}. 

The function $L^{(\alpha)}$ is defined as
\begin{equation}
L^{(\alpha)} = L_{\alpha} + L_{\tau} + L_{\mu} 
+ L_{e} + \eta,
\end{equation}
where $\eta$ is approximately equal to the baryon to photon ratio, and the
$L_{\nu}$'s are the lepton asymmetries which are given by
\begin{equation}
L_{\nu_{\alpha}}= \frac{n_{\nu_\alpha}-n_{\bar{\nu}_\alpha}}{n_\gamma},
\end{equation}
where $n_{\nu}$ and $n_{\gamma}$ are the neutrino and 
photon number densities, respectively. 

The neutrinos have a distribution of momenta and only neutrinos 
with momentum of a certain value will be resonant at a given time.  
The momentum of the resonance will move through the neutrino 
distribution as the temperature decreases, creating a
neutrino asymmetry at the particular value of the 
resonance momentum at that
temperature. The evolving neutrino asymmetry also non-linearly 
affects the evolution of the resonance momentum.  The 
asymmetry is distributed across the momentum
distribution as weak interactions keep the distributions 
in thermal equilibrium.

The resonance occurs for 
\begin{equation}
(b\pm a) = \cos2\theta_0\simeq 1,
\end{equation} 
for $\sin^22\theta_0\ll1$.  The $a$ and $b$ terms dominate 
this condition at different temperatures.  We may obtain 
a qualitative understanding of the behaviour
by neglecting the thermal spread of momentum and 
using the average value $\langle
p\rangle \simeq 3.15T$.  At high temperatures the 
$\langle b \rangle$ term dominates, and in
\cite{rr1} it was found that for $\langle b \rangle<1$ 
oscillations create lepton number while for
$\langle b \rangle >1$ the oscillations are lepton 
number destroying. The dependence
of $\langle b \rangle$ on
temperature is $\langle b \rangle \sim T^6 $, so when 
the temperature drops to where
$\langle b \rangle \sim 1$, which we denote by $T_C$, 
the oscillations creating lepton asymmetry
dominate and lead to an exponential growth in lepton 
number.  
In Ref.\cite{rr1} it was shown that for $\delta m^2<0$ 
(and $|\delta m^2|\agt 10^{-4}\text{eV}^2$) large lepton 
asymmetries are created provided that 
\begin{equation}
\sin^2 2\theta_0 
\agt 5 \times 10^{-10}\left(
{\text{eV}^2 \over |\delta m^2|}\right)^{1/6}.
\end{equation}
This result is independent of the initial value of the asymmetry, 
provided that $|L_{\alpha}|\alt 10^{-5}$ \cite{rr1}.
For $\delta m^2$ in the
range $1-10\text{eV}^2$, we have $T_C\simeq 16 - 25\text{MeV}$, 
obtained from the condition $\langle b \rangle =1$.  
The rapid growth drops off as the temperature
decreases and when the temperature has decreased to below about 
$T_C/2$, the $\langle a \rangle$ term
has become larger and we can now neglect 
the $\langle b \rangle$ term so that the resonance condition
is approximately given by \cite{rr,rr1}
\begin{equation}
\langle a \rangle \simeq \cos2\theta_0\simeq 1.
\end{equation} 
Collisions and oscillations both affect the 
generation of lepton asymmetries.  At
high temperatures $(T\agt T_C/2)$ collisions 
affect neutrinos to a 
greater extent than oscillations, and these in turn
affect lepton number because the $\nu$'s and $\bar{\nu}$'s 
interact at different rates in a CP asymmetric background 
medium. (A CP asymmetric medium is required by
the cosmological baryon-antibaryon asymmetry.)
At lower temperatures, the collision
rates are not as rapid and oscillations become coherent, 
so that oscillations take
over from collisions to control the growth of lepton number.  
Thus, for $T\alt T_C/2$, MSW \cite{kuo} transitions become important. 

We have two oscillation resonances to consider, 
$\nu_{\tau,\mu}-\nu_s$, which will generate a
large $L_{\tau,\mu}$ asymmetry and $\nu_{\tau,\mu}-\nu_e$, 
which will then transform
some of the asymmetry into the $\nu_e$ sector 
to produce $L_e$.  Oscillations between
$ \nu_{\tau}$ and $\nu_{\mu}$ or $\nu_e$ and $\nu_s$ 
may be neglected as they have
much smaller mass-squared differences. 

The asymmetries are calculated, starting from an initial 
temperature of $T=T_C/2$
where $|L_{\tau}|\ll 1$ and T is low enough for oscillations 
and MSW transitions to
dominate over collisions.  The initial number density 
of sterile neutrinos is taken
to be negligible (which will occur for a range of parameters) 
in accord with the results of the numerical calculation in
Ref.\cite{rr1} for the initial creation of lepton number at $T=T_C$.  
We study how
the asymmetries evolve down to temperatures where they become 
important for BBN.
Note that for the moment we will assume that $L_\tau > 0$
for definiteness.

At $T\agt T_C$ where the $b$ term dominates, 
$\nu_{\tau}\rightarrow\nu_s$ and
$\bar{\nu}_{\tau}\rightarrow\bar{\nu}_s$ resonances both 
occur at the same momentum, but as the temperature falls to 
$T\simeq T_C/2$ the $\nu_{\tau}\rightarrow\nu_s$ and
$\bar{\nu}_{\tau}\rightarrow\bar{\nu}_s$ oscillations no 
longer have the same resonance momenta.  
This allows the asymmetry to grow very large because when most of
the $\bar{\nu}$'s have approximately the right momentum 
to be near a resonance, the
bulk of the $\nu$'s are far away from 
a resonance.  It was found in \cite{rr} that
if the created $L_{\alpha}$ is positive the 
antineutrino resonance momentum
has generally low values $p_{res}/T\sim 0.2-0.8$ at $T \simeq T_C/2$ 
while the neutrino resonance momentum is toward the tail 
of the neutrino momentum 
distribution (see Fig.2 of Ref\cite{rr} for an illustration).  
Hence the neutrino resonance is neglected and 
we will only be concerned with the antineutrino resonance. 
     
We will neglect the $\nu_{\tau}-\nu_{\mu}$ mass difference 
and thus make the approximation
that the $\nu_{\tau}-\nu_s$ and $\nu_{\mu}-\nu_s$ oscillations 
have exactly the same
resonance momentum, $p_1$, and the $\nu_{\tau,\mu}-\nu_e$ 
resonances have momentum $
p_2$.  These are obtained from the resonance condition 
\begin{equation}
\label{eq:res}
a \simeq \cos2\theta_0\simeq 1,
\end{equation}
where $a$ is given by eq.(\ref{eq:a}).  The resonance 
momenta are given by
\begin{eqnarray}
p_1 \simeq -
\frac{\pi^2 \delta m^2}{4\sqrt{2}\zeta (3)G_F}\frac{1}{T^3L^{(\tau)}}, 
\nonumber \\
p_2 \simeq -
\frac{\pi^2 \delta m^2}
{4\sqrt{2}\zeta (3)G_F}\frac{1}{T^3(L^{(\tau)}-L^{(e)})}.
\end{eqnarray}

Since $\nu_{\tau}$ and $\nu_{\mu}$ are nearly degenerate 
and their matter effects are
identical, the calculation is symmetric with 
respect to $\nu_{\mu}$ and $\nu_{\tau}$,
so $L_{\tau}=L_{\mu}$.  We therefore find that
\begin{eqnarray}
L^{(\tau)}=3L_{\tau}+L_e \simeq 3L_{\tau}, \nonumber \\
L^{(\tau)}-L^{(e)}=L_{\tau}-L_e \simeq L_{\tau}.
\end{eqnarray}
This means that the $\nu_{\tau,\mu}-\nu_e$ resonance occurs 
for a momentum which is
approximately a factor of three higher than for the 
$\nu_{\tau,\mu}-\nu_s$ resonance,
\begin{equation}
\label{eq:p3}
p_2\simeq 3p_1.
\end{equation}
Because the $\nu_{\tau}-\nu_s$ and $\nu_{\mu}-\nu_s$ 
oscillations have the same resonance 
momenta, instead of having two separate resonances, 
the $\nu_{\tau}-\nu_s$ and 
$\nu_{\mu}-\nu_s$ resonances completely overlap.  
Therefore we must consider the 
three-flavour system consisting of $\nu_{\tau}$, $\nu_{\mu}$ and 
$\nu_{s }$. To take into account that there is only one 
sterile neutrino state for both 
$\nu_{\tau}$ and $\nu_{\mu}$ to couple to, 
we may form two linear combinations of 
$\nu_{\tau}$ and $\nu_{\mu}$.  One of 
these oscillates with $\nu_s$, and undergoes 
MSW conversion at the resonance, while the 
other $\nu_{\tau}$/$\nu_{\mu}$ linear 
combination does not oscillate with $\nu_s$.  This state is decoupled, 
and is not converted into the sterile state at the resonance.  
Our equations include factors of $\frac{1}{2}$ 
to account for this situation.  The mass eigenstate 
which is dominantly a $\nu_s$ at high density, becomes a 
$\nu_{\tau}$/$\nu_{\mu}$ linear combination at low density, 
so that a $\nu_s$ which 
goes through the resonance has a 50\% chance of being 
converted to $\nu_{\tau}$ 
and a 50\% chance of being converted to $\nu_{\mu}$.  Similarly, 
since one linear combination of $\nu_{\tau}$ and 
$\nu_{\mu}$ only will be converted to a $\nu_s$ upon 
passing through the resonance, both $\nu_{\tau}$ 
or $\nu_{\mu}$ have only a 50\% 
probability of being converted to the sterile state.

Naively one might believe that by making $\nu_{\mu}$ 
and $\nu_{\tau}$ degenerate, so
as to have two heavier neutrinos oscillating with 
$\nu_s$ instead of only one heavier
neutrino oscillating with the sterile neutrino, 
that twice the electron-neutrino 
asymmetry would be produced.  However, this will 
not occur, because there is only one sterile
eigenstate for the two linear combinations 
of $\nu_{\tau}$ and $\nu_{\mu}$ to
oscillate with.  Due to the factors of $\frac{1}{2}$ 
arising from this three-flavour
nature, we would not expect the electron-neutrino 
asymmetry created to be any larger
than for the model of Ref.\cite{rr}, where only a heavy 
$\nu_{\tau}$ oscillates with
$\nu_s$ (in fact it turns out to be smaller -see later).

Numerically integrating the quantum kinetic equations, we find 
that the adiabatic limit of the MSW 
effect holds provided
\begin{equation} 
\label{eq:robert}
\sin^2 2\theta_0 \agt 10^{-9} - 10^{-10}
\end{equation}  
for $\delta m^2 \sim 10\text{eV}^2$.
The adiabatic limit implies all neutrinos that pass
through the resonance will undergo MSW conversion.  
This means the lepton number
created is basically equal to the number of 
antineutrinos minus the number of sterile
antineutrinos that pass through the resonance, and 
the growth of lepton number is
governed by how quickly the resonance momentum moves 
through the neutrino momentum distribution. 

The rate of change of the lepton numbers is therefore given by 
\begin{eqnarray}
\frac{dL_{\nu_{\tau}}}{dT} = 
-X_1\left|\frac{d}{dT}\left(\frac{p_1}{T}\right)\right| 
- X_2\left|\frac{d}{dT}\left(\frac{p_2}{T}\right)\right|, \label{eq:dL3} \\
\frac{dL_{\nu_e}}{dT} = 
2X_2\left|\frac{d}{dT}\left(\frac{p_2}{T}\right)\right|,
\label{eq:dL1}
\end{eqnarray}
where $X_1$($X_2$) expresses the difference between 
the number of $\bar{\nu}_{\tau}$ and
$\bar{\nu}_s$($\bar{\nu}_e$) with the right momentum to pass 
through the resonance and is given by
\begin{eqnarray}
\label{eq:xx}
X_1 = \left.\frac{1}{2}\frac{T}{n_\gamma}
(N_{\bar{\nu}_\tau} - N_{\bar{\nu}_s})\right|_{p=p_1}, \nonumber \\
X_2 = \left.\frac{1}{2}\frac{T}{n_\gamma}(N_{\bar{\nu}_\tau} 
- N_{\bar{\nu}_e})\right|_{p=p_2}.
\end{eqnarray}
The factor of 2 in Eq.(\ref{eq:dL1}) and the factors of $\frac{1}{2}$ in
Eqns.(\ref{eq:xx}) arise 
from the three-flavour nature of the problem as discussed above. 

The N's are the momentum distributions of the neutrinos, so that
\begin{equation}
n=\int Ndp,
\end{equation}
which in thermal equilibrium are just given by a Fermi distribution
\begin{equation}
N_{\bar{\nu}}=
\frac{1}{2\pi^2}\frac{p^2}{1+\exp{\left(\frac{p+\mu_{\bar{\nu}}}{T}\right)}},
\label{nic}
\end{equation}
where $\mu_{\bar{\nu}}$ is the antineutrino chemical potential.  The
$T\left|\frac{d}{dT}\left(\frac{p_{res}}{T}\right)\right|$ terms 
describe how quickly
the resonance momenta move, and are expressed in terms of the variable
$\frac{p_{res}}{T}$ because we want to distinguish the roles of 
expansion and oscillations, and unlike $p$, the 
quantity $\frac{p}{T}$ does not change due to the
direct effect of the expansion of the universe.  Evaluating 
the derivatives, we obtain
\begin{eqnarray}
\frac{dL_{\nu_{e}}}{dT} = \frac{2DA + CE}{A + E(A-B)}, \label{eq:L3} \\
\frac{dL_{\nu_{\tau}}}{dT} = \frac{C}{A} + \frac{B}{A}\frac{dL_{\nu_{e}}}{dT},
\label{eq:L1}
\end{eqnarray}
where
\begin{eqnarray}
A = - \left( 1 + \frac{3X_1p_1}{TL^{(\tau)}}\right), \hspace{5mm} 
B = \frac{1}{2} + \frac{X_1p_1}{TL^{(\tau)}},    \nonumber  \\
C = \frac{4X_1p_1}{T^2}, \hspace{5mm} D = \frac{4X_2p_2}{T^2}, \hspace{10mm}
\nonumber \\
E = \frac{2X_2p_2}{T}\frac{1}{L^{(\tau)}-L^{(e)}}.\hspace{15mm}
\end{eqnarray}
In deriving eqns.(\ref{eq:L3},\ref{eq:L1}) we have used that
$\frac{d}{dT}\left(\frac{p_{1,2}}{T}\right) < 0$, because 
the resonances (which start at low values of approximately 
$p_1/T\sim0.2-0.8$ and $p_2/T \simeq 3p_1/T$) move
through the momentum distributions to higher values as 
the temperature decreases. 

Numerically integrating eqns.(\ref{eq:L3},\ref{eq:L1}), 
we find that the final lepton asymmetries generated are
\begin{equation}
L_{\mu}/h = L_{\tau}/h \simeq 0.16, \;\;\;\; L_e/h \simeq 6.7\times10^{-3},
\end{equation}
for  $\delta m^2$ in the range 6-8eV$^2$, where 
$h=T_{\nu}^3/T^3_{\gamma}$.

The $L_e$ asymmetry then affects BBN rates via 
the modification of the neutrino momentum distributions.  
Weak interactions keep the neutrino distribution in thermal equilibrium 
for $T \agt 1\text{MeV}$, however for $\delta m^2 \sim 10\text{eV}^2$
the neutrino asymmetries only become large at temperatures 
where the the distributions are beginning to go out of equilibrium. 
Because the neutrino distributions are not completely thermalised, we 
cannot simply describe them in terms of chemical potentials, so for 
this reason we keep track of the neutrino number densities numerically 
in momentum cells.  We assume complete MSW conversion at the resonance, which 
is a good approximation for the large range of parameters given 
by Eq.(\ref{eq:robert}), so that numbers densities of sterile and 
active neutrinos are interchanged at the resonance momentum such that
\begin{eqnarray}
N_{\bar\nu_{s}}(p_1) \rightarrow \frac{1}{2}(N_{\bar\nu_{\tau}}(p_1)+N_{\bar\nu_{\mu}})(p_1) = N_{\bar\nu_{\tau,\mu}}(p_1) \nonumber \\
N_{\bar\nu_{\tau,\mu}}(p_1) \rightarrow \frac{1}{2}(N_{\bar\nu_s}(p_1)+N_{\bar\nu_{\tau,\mu}}(p_1)),
\end{eqnarray}  
and similarly for the $\bar\nu_{\tau,\mu}-\bar\nu_e$ resonance. 
Cells are refilled according to the interaction rates
\begin{equation}
\frac{d}{dt}\left( \frac{N_{actual}}{N_0} \right) = \Gamma(p) 
\left( \frac{N_{equilib}}{N_0}-\frac{N_{actual}}{N_0} \right), 
\label{eq:refill}
\end{equation}
where 
\begin{equation}
N_0=\frac{1}{2\pi^2}\frac{p^2}{1+\exp{\left(\frac{p}{T}\right)}}.
\end{equation}
and $N_{equilib}$ is given by the equilibrium distribution
Eq.(\ref{nic}). The chemical potential is obtained from
the lepton number via the equation
\begin{equation}
L_{\alpha} \simeq -{1 \over 24\zeta(3)}\left[
\pi^2(\tilde{\mu}_{\nu} - \tilde{\mu}_{\bar \nu})
-6(\tilde{\mu}_{\nu}^2 - \tilde{\mu}^2_{\bar \nu})
\ln 2 + (\tilde{\mu}^3_{\nu} - 
\tilde{\mu}^3_{\bar \nu})
\right]
\label{robo}
\end{equation}
where $\tilde{\mu}_i \equiv \mu_i/T$.
Equation(\ref{robo}) is an exact equation for $\tilde{\mu}_{\nu}
= -\tilde{\mu}_{\bar \nu}$, otherwise it holds to
a good approximation provided that
$\tilde{\mu}_i \alt 1$.
For $\delta m^2 \sim 8\ \text{eV}^2$ significant lepton number
(here significant means larger than about $0.01$)
is not created until $T \alt 2-3 \text{MeV}$.
At these temperatures,
the neutrinos and anti-neutrinos have already chemically decoupled.
(Note that $T^e_{dec} \simeq 3$ MeV and $T^{\mu,\tau}_{dec}
\simeq 5$ MeV are the chemical decoupling temperatures).
Because of this, $\tilde{\mu}_{\nu} \simeq 0$,
while the anti-neutrino chemical potential $\tilde{\mu}_{\bar \nu}$ 
continues increasing as per eq.(\ref{robo}).

The interaction rate $\Gamma(p)$ in eq.(\ref{eq:refill}) is 
\begin{equation}
\Gamma(p)\simeq \frac{\langle \Gamma\rangle p}{\langle p \rangle},
\end{equation}
where $\langle \Gamma \rangle$ is the thermal average of the total neutrino collision rate 
given by \cite{enqvist}

\begin{eqnarray}
\langle\Gamma_{\nu_e}\rangle=4.0G_F^2T^5,  \nonumber \\
\langle\Gamma_{\nu_{\tau,\mu}}\rangle=2.9G_F^2T^5.
\end{eqnarray} 

Integrating the rate equations 
for the processes given in
eq.(\ref{eq:rates}), using the modified neutrino momentum 
distributions ($N_{actual}(p)$), the change
in the neutron to proton ratio and hence $Y_P$ is determined.  
We find the size of
the effect in this case is given by $\delta Y_P\simeq-0.0023$ which 
corresponds to a
change in the effective number of neutrinos 
of $\delta N_{\nu}^{eff}\simeq-0.19$. 

Excitation of the sterile neutrino, and the modified momentum 
distribution, will
change the energy density of the universe, 
and this will also change
the effective number of neutrinos. 
The number density and average energy of the neutrinos 
were calculated from the
modified momentum distributions.  For $\delta m^2$ 
in the range $ 6-8\text{eV}^2$ we get
\begin{eqnarray}
\label{eq:n}
\frac{n_{\bar{\nu}_{e}}}{n_0}=0.99,  \;\;  
\frac{\langle E_{\bar{\nu}_{e}}\rangle}{\langle p \rangle}=0.99, 
\nonumber \hspace{15mm} \\
\frac{n_{\bar{\nu}_{\tau}}}{n_0}=\frac{n_{\bar{\nu}_{\mu}}}{n_0}=0.53, 
\;\; \frac{\langle E_{\bar{\nu}_{\tau}}\rangle}{\langle p \rangle}=
\frac{\langle E_{\bar{\nu}_{\mu}}\rangle}{\langle p \rangle}=0.91, \nonumber \\
\frac{n_{\bar{\nu}_{s}}}{n_0}=0.91, 
\;\; \frac{\langle E_{\bar{\nu}_{s}}\rangle}{\langle p \rangle}=0.97, \hspace{15mm}
\end{eqnarray}
where $n_0$ is the number density with zero chemical 
potential and $\langle p \rangle=3.15T$ is the 
average momentum for a Fermi-Dirac distribution.  (The chemical
potentials for the neutrinos are 
very small and so the distributions are not
significantly modified.)  These changes 
in the energy density correspond to $\delta
N_{\nu}^{eff}\simeq-0.09$.  As we 
have an ambiguity concerning the sign of the lepton
asymmetries and hence also the sign of $\delta N_{\nu}^{eff}$ 
due to modification of
the nuclear reaction rates, we have two possible 
values for the overall change in
the effective number of neutrinos.

In fig.(\ref{dn}) $\delta N_{\nu}^{eff}$ is plotted as a 
function of the mass squared
difference.  For the mass range  $\delta m^2
\simeq 6-8\text{eV}^2$, we find: 
\begin{eqnarray}
\delta N_{\nu}^{eff}\simeq-0.3,  \;\; \text{positive asymmetry}, \nonumber \\
\delta N_{\nu}^{eff}\simeq+0.1,   \;\; \text{negative asymmetry},
\end{eqnarray}
This remains approximately constant to above where the 
NOMAD experiment cuts off the
LSND data at $\delta m^2 \simeq 10\text{eV}^2$ \cite{nomad}, 
while for small $\delta m^2$ we have 
$\delta N_{\nu}^{eff}\rightarrow 0$.

The $L_e$ asymmetry and hence the $\delta N_{\nu}^{eff}$ 
obtained is smaller than
that obtained for the model of Ref.\cite{rr}.  
The $L_{\tau,\mu}$ asymmetry created in the present model is 
roughly half as big as the $L_{\tau}$ asymmetry obtained 
in Ref.\cite{rr}, because only one of the the two linear 
combinations of $\nu_{\mu}$ and $\nu_{\tau}$ oscillates 
with $\nu_s$, as discussed above.  
Naively one
might expect that with both $\nu_{\tau}-\nu_e$ and 
$\nu_{\mu}-\nu_e$ oscillations
producing $L_e$ that an effect of comparable size 
to Ref.\cite{rr} would be obtained,
however, this does not turn out to be correct.  
The crucial difference between the
two cases is the separation of the two resonances. The 
separation of the resonances
is given by Eq.(\ref{eq:p3}), 
and whereas we have $p_2\simeq3p_1$, in the previous
case the resonances were closer together 
with $p_2\simeq2p_1$.  This is important
because the resonance which creates $L_e$ is at 
a higher momentum than the resonance
which creates $L_{\tau}$, so that by the 
time $L_{\tau}$ has grown sufficiently large
to start producing significant $L_e$, 
the $p_2$ resonance has already moved a
considerable way through the neutrino momentum distribution.  
For the present model, $p_2/T$ has already moved past 
the average value of $p/T\simeq3.15$ and is out toward
the tail of the distribution, where the number 
density is much lower, so that the
$L_{\tau}$ asymmetry cannot be transferred to $L_e$ as efficiently. 

\section{Conclusion}

We conclude that the neutrino oscillations in the 
four neutrino model of Ref.\cite{model} can 
modify the effective number of neutrinos during nucleosynthesis and
give $N^{eff}_{\nu} < 3$. However, it turns out that the
effect is relatively small, $\delta N^{eff}_{\nu} \simeq -0.3$ 
(if $L_e > 0$).
If future measurements confirm the low D/H result as well as
the $Y_P$ value used in Ref.\cite{hata},
then this model will become cosmologically disfavoured.  
On the other hand, if future
measurements favour $N_{\nu}^{eff}\simeq 3$ then 
this model will remain viable. 

There is also the interesting possibility that 
the neutrino mass favoured for
dark matter \cite{primack} will have to 
be recalculated. 
In fact, from Eq.(\ref{eq:n}) the number of heavy neutrinos is actually 
about 1.5 because about $50\%$ of the $\bar{\nu}_{\tau}$ and 
$\bar{\nu}_{\mu}$ have been converted into light $\bar{\nu}_s$.  
This would suggest that the optimal mass for hot+cold dark matter 
senarios is somewhat heavier, about $m_{\nu_{\tau}}=m_{\nu_{\mu}}=3.1\text{eV}$ 
rather than $2.4\text{eV}$, for this particular model.  This implies that 
the $\delta m^2$ relavant for LSND
is about $\text{9eV}^2$ rather than $\text{6eV}^2$ \cite{footnote}.

\acknowledgments{We thank David Caldwell for prompting 
this investigation.  RF and RRV are supported by the 
Australian Research Council.  NFB is supported by the 
Commonwealth of Australia and the University of 
Melbourne.  NFB thanks Y.Y.Y. Wong for useful discussions.}

\begin{figure}

\caption{Change in the effective number of neutrinos as a function of the
mass-squared difference.  The solid 
line corresponds to a positive electron neutrino
asymmetry, and the dashed line a negative asymmetry.}

\newpage
\epsfig{file=dN.epsi,width=15cm}
\label{dn}
\end{figure}


\begin{references}
\bibitem{footnote}
Of course this assumes that $\sin^2 2\theta_0\agt 3\times10^{-10}$.  
If the mixing angle is smaller than this, lepton number is not generated 
by oscillations and consequently $m_{\nu_{\tau}}=m_{\nu_{\mu}}=2.4\text{eV}$ 
would remain as the favoured neutrino mass for hot+cold dark matter scenarios.
\end{references}
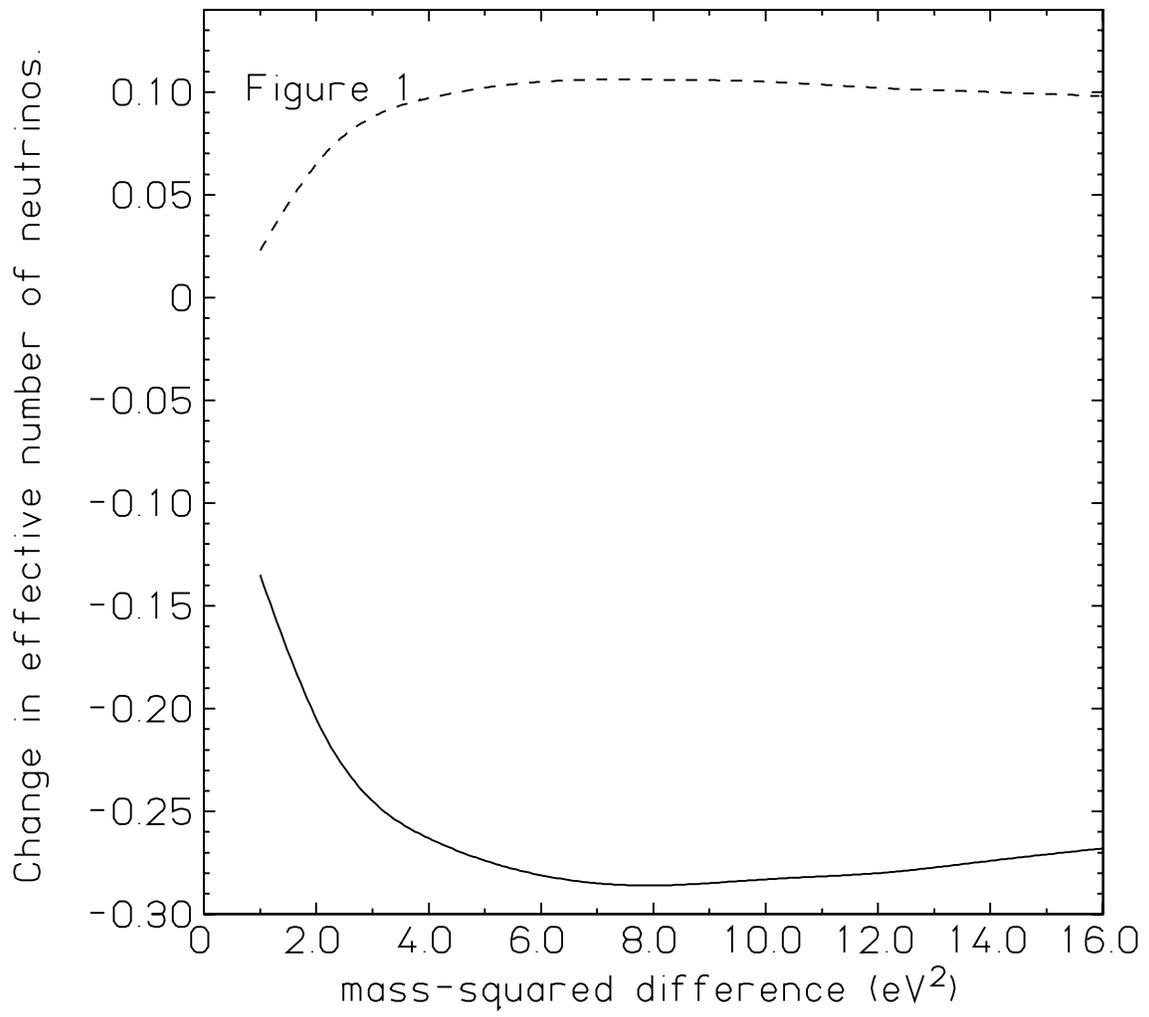
\end{document}